\documentclass[a4paper]{article}
\usepackage{icrc2013}
\usepackage[english]{babel}

\usepackage{amssymb}
\title{SST-GATE: A dual mirror telescope for the Cherenkov Telescope Array}

\shorttitle{The SST-GATE telescope prototype}

\authors{
Zech, A.$^1$,
Amans, J.-P.$^2$,
Blake, S.$^3$,
Boisson, C.$^1$,
Costille, C.$^2$,
De-Frondat, F.$^2$,
Dournaux, J.-L.$^2$, 
Dumas, D.$^2$, 
Fasola, G.$^2$,
Greenshaw, T.$^4$,
Hervet, O.$^1$,
Huet, J.-M.$^2$,
Laporte, P.$^2$,
Rulten, C.$^1$, 
Savoie, D.$^2$,
Sayede, F.$^2$,
Schmoll, J.$^3$,
Sol, H.$^1$ 
for The CTA Consortium }

\afiliations{
$^1$ LUTH, Observatoire de Paris, CNRS, Universit\'{e} Paris Diderot ; 5 Place Jules Janssen, 92190 Meudon, France\\
$^2$ GEPI, Observatoire de Paris, CNRS, Universit\'{e} Paris Diderot, France\\
$^3$ Durham University, CfAI, NetPark Research Institute, UK\\
$^4$ University of Liverpool, UK
}

\email{andreas.zech@obspm.fr}

\abstract{
The Cherenkov Telescope Array (CTA) will be the world's first open observatory for very high energy $\gamma$-rays. Around a hundred telescopes of different sizes will be used to detect the Cherenkov light that results from $\gamma$-ray induced air showers in the atmosphere. Amongst them, a large number of Small Size Telescopes (SST), with a diameter of about  4\,m, will assure an unprecedented coverage of the high energy end of the electromagnetic spectrum (above $\sim$1\,TeV to beyond 100\,TeV) and will open up a new window on the non-thermal sky. \\
Several concepts for the SST design are currently being investigated with the aim of combining a large field of view ($\sim$9 degrees) with a good resolution of the shower images, as well as minimizing costs. These include a Davies-Cotton configuration with a Geiger-mode avalanche photodiode (GAPD) based camera, as pioneered by FACT, 
and a novel and as yet untested design based on the Schwarzschild-Couder configuration, which uses a secondary mirror to reduce the plate-scale and to allow for a wide field of view with a light-weight camera, e.g. using GAPDs or multi-anode photomultipliers.\\
One objective of the GATE (Gamma-ray Telescope Elements) programme is to build one of the first Schwarzschild-Couder prototypes and to evaluate its performance. The construction of the SST-GATE prototype on the campus of the Paris Observatory in Meudon is under way. We report on the current status of the project and provide details of the opto-mechanical design of the prototype, the development of its control software, and simulations of its expected performance.
}
 

\begin{document}
\maketitle

\section{Introduction}

The CTA~\cite{ach2013} will be the first open observatory for the detection of $\gamma$-rays at very high energies (VHE; $\gtrsim$30\,GeV) and will be driven by observation proposals from the scientific community. The project is currently in its prototyping phase and construction of the first telescopes is expected to commence in 2015. A wide energy range (spanning five orders of magnitude) will be achieved with telescopes of three different types and sizes: energies of about 100 GeV up to tens of TeV will be covered with mid size telescopes (MSTs; 12\,m diameter), while large size telescopes (LSTs; 24\,m diameter) will allow for the imaging of weak air showers from $\gamma$-rays in the
10 - 100 GeV range, and a large number of small size telescopes (SSTs), spread out over a wide area, will provide access to very luminous, but sparse, $\gamma$-ray showers up to and beyond 100 TeV. 

\begin{figure}[!h]
  \centering
\includegraphics[width=0.86\columnwidth]{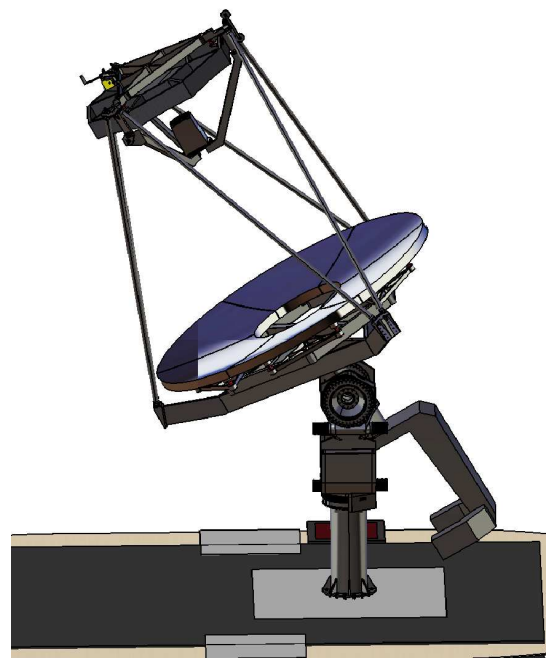}
  \caption{\small{Computer model (using the CATIA software) of the SST-GATE telescope. The camera is located between the primary and secondary mirror.}}
  \label{fig:sstgate}
\end{figure} 

The SST component of CTA will thus extend the energy reach beyond the limits of current Imaging Air Cherenkov Telescope (IACT) arrays to explore the high energy end of Galactic and extragalactic sources and probe the limits of particle acceleration in TeV emitters. To achieve this objective, a sufficiently large number of SSTs (on the order of 70) are needed to cover a surface area of $\gtrsim$3\,km$^2$. The cost for a single telescope thus needs to be reduced as much as is feasible. A small effective mirror area of at least 
5\,m$^2$ is considered sufficient for the SSTs, given the bright signal from air showers in the targeted energy range. On the other hand, the viewing angle of the telescopes needs to be
large, on the order of 10$^{\circ}$, to allow the imaging of showers that are far away from each single telescope, and thus imaged under larger field angles. This will permit
the installation of a spread-out SST array with a large inter-telescope spacing (up to $\sim$300 m). The optimal configuration of such an array is currently under
study with large scale Monte Carlo simulations of air showers and of the instrumental response of different telescope designs~\cite{ctamc}.

Three different designs are currently being considered by the CTA consortium: one using single-mirror Davies-Cotton (DC) optics with a GAPD based camera~\cite{Nie2013}, similar to the FACT telescope \cite{And2013}, and two others, the ASTRI \cite{astri} and SST-GATE projects, based on Schwarzschild-Couder (SC) optics \cite{Vas2007}. The latter is an aplanatic, anastigmatic design developed by Andr\'{e} Couder \footnote{who happens to be a former director of the Optical Laboratory at the Paris Observatory} in the 1920s as a variation of Schwarzschild's optical systems. In the SC telescope, the focal plane is located in-between two aspherical mirrors, close to the secondary mirror (cf. Fig.~\ref{fig:sstgate}). The plate-scale is much smaller than for the DC design, permitting the use of a small, light-weight camera, while still providing a wide Field of View (FoV). This will provide the required performance, with reduced costs for the camera and telescope structure. The challenge lies in the fact that such a telescope has never been built before and that practical solutions for the non-spherical mirrors still need to be tested. 

\section{A Brief Description of SST-GATE}

SST-GATE represents a light-weight, cost-efficient SC telescope that will be easy to install and maintain on the CTA site. This section provides
a short overview of the optical system, the mechanical structure and the telescope control software. Table~\ref{tab:specs} lists some of the main characteristics
of the SST-GATE design.

\begin{table}[h]
\begin{center}
\begin{tabular}{|l|c|}
\hline 
diameter primary mirror  & 4\,m   \\ \hline
diameter secondary mirror  & 2\,m \\ \hline
effective collecting area (on-axis) & $\sim$8.2\,m$^2$ \\ \hline
field of view & 9$^\circ$ \\ \hline
plate-scale & 39.6\,mm/$^\circ$ \\ \hline
on-axis PSF & $\sim$0.05$^\circ$ \\ \hline
total weight & 7.8\,t \\ \hline
\end{tabular}
\caption{Principal characteristics of the SST-GATE design. The effective collection area is corrected for obscuration by the secondary mirror, camera and masts.
The PSF is for a point-source at a distance of 5\,km (cf. Section \ref{sec:sim}). }
\label{tab:specs}
\end{center}
\end{table}

%\vspace{0.1cm}

\subsection{Optical System}

\begin{figure}[h]
  \centering
\includegraphics[width=\columnwidth]{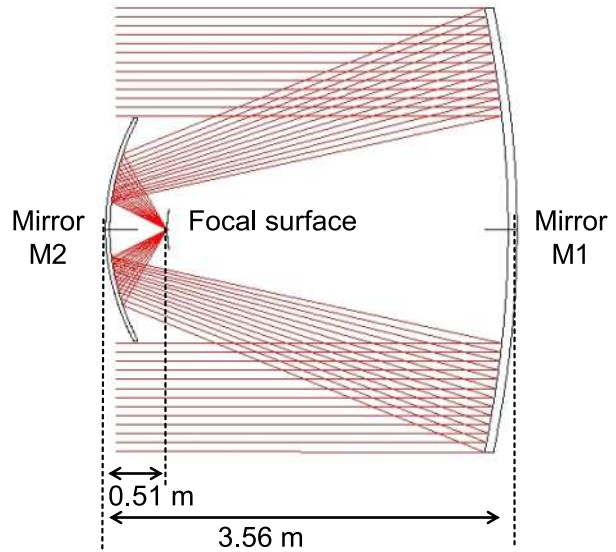}
  \caption{\small{Light path and dimensions of the optical structure.}}
  \label{fig:optics}
\end{figure}

A schema of the optical system is shown in Fig.~\ref{fig:optics}. The mirror surfaces are described by 16th order polynomials to minimize the size of the PSF over
the whole field of view. Obscuration by the secondary mirror (M2) leads to a reduction of the effective mirror surface of the primary (M1) by 25\% for on-axis light rays. The center of the primary mirror surface, which is obscured for all field angles, has a hole of 65\,cm radius, which allows the integration of a laser for alignment purposes.

For the prototype, M1 will be made of six equally sized aluminium segments that will be aligned with the help of three actuators per segment. 
The smaller M2 is foreseen to be a monolithic mirror made of glass, stiffened with a honeycomb structure, and fabricated in a cold slumping process. 
It will equally be attached to the dish structure with three actuators. 
Both mirrors can thus be moved along the optical axis within the range of the actuators to permit an adjustment of the focal plane onto the detector plane.

The curved focal plane will be equipped with a specially designed camera comprising GAPDs or multi-anode photomultipliers.
The SST camera is being developed by CTA groups from the UK, USA and Japan, known as the CHEC project~\cite{chec}. The pixel size will be 6$\times$6\,mm$^2$ and the total camera surface will have a diameter of 36\,cm. For optimum resolution, the PSF should be contained within a single pixel, corresponding to an angle of 0.15$^{\circ}$ out to the edge of the field of view.

The alignment procedure foresees the installation of a laser behind the center of M1 to define the optical axis of the telescope, serving as a reference for the alignment of the mechanical axis. The M2 mirror and the segments of M1 will be aligned with the help of an artificial light source. A pointing monitor, using the position
of stars as references, is currently being developed by the Dutch CTA group. We are considering the possibility of attaching this monitor back-to-back with M2 and aligning it with the telescope axis. The absolute pointing position of the SST-GATE telescope, after post-processing of the data, will be below the required 7 arc seconds RMS.

%\vspace{0.1cm}

\subsection{Mechanical Structure}

The design and optimization of the mechanical telescope structure is the main focus of the SST-GATE project. Detailed Finite Element Analyses (FEA) and modal analyses were carried out to optimize the structure within the given specifications for the CTA SSTs. The influence of wind load, seismic and temperature effects were evaluated to be compliant with the requirements of the potential CTA sites and of the site in Meudon. 

The telescope structure is made up of several subsystems. The alt-azimuthal Subsystem (AAS), made of steel, consists of a tower with a fork mount, and the azimuth and elevation drives, based on electric torque motors. The Mast and Truss Structure, made of steel truss tubes in a Serrurier configuration, connects the AAS to M2 and to the camera, while M1 is supported by a dedicated dish linked directly to the AAS. This decoupling of the dish of M1 from the remaining optical structure decreases deformations of the M1 dish during operation and improves the off-axis performance of the telescope.

The camera is supported by three arms connected to a triangular support behind M2, thus guaranteeing a high precision in the distance between camera and M2, 
which is critical for the image quality (the tolerance for a displacement between M2 and the camera being of the order of 1\,mm). The camera can be dismounted by unlocking the upper arm and rotating it downwards to a height of about 1\,m above ground or
completely onto the ground with the help of a simple pulley system. This provides great ease of access for maintenance operations and for the exchange of cameras. 

The telescope is balanced with an adaptable counterweight. The complete structure weighs only about 7.8 tons. This light weight is a strong advantage of our design, since it reduces the requirements on the foundation and eases the transport and assembly of its parts. 

On the site in Meudon, the telescope prototype will be protected with a retractable shelter, already installed, in the form of a dome with a PVC membrane, in order to decrease the number of re-coating operations for the mirrors and to provide protection from animals, vandalism and the elements. The shelter is sufficiently large to permit assembly and maintenance operations while the cover is closed.

\vspace*{0.4cm}

\subsection{Telescope Control Software}

The control software for SST-GATE is divided into two main parts: the telescope software, implemented in a CompactRIO from National Instruments, and the remote software, implemented on a workstation. The telescope software is built to be operated either in stand-alone mode or --- thanks to the OPCUA server used as a gateway with the remote software and other possible clients --- in an array. The remote software allows high level orders to be given to the telescope and provides a user-friendly view of the telescope parameters. 

Since the communication architecture is based entirely on Ethernet connections, it can be run on a workstation in the control room on the Meudon site or from anywhere else via an internet connection. Optical fibres are used only to read out the data from the camera used for observations.

The software controls the telescope drives for pointing and tracking, monitors the telescope status, interacts with actuators and sensors for the optical alignment, and 
communicates with the focal plane instrumentation. To provide the required pointing and tracking precision for the prototype telescope, the main computer clock will be synchronized via a daemon with a secondary (stratum 2) NTP time server provided by the Paris Observatory.

The programming language for the remote software is LabVIEW. A Graphical User Interface will give access to predefined functions of the telescope control even to 
non-expert users, thus allowing the use of the prototype for student training. 

\vspace{0.1cm}

\section{Simulations of the Optical Performance}
\label{sec:sim}

The optical performance of the SST-GATE telescope --- its expected throughput and PSF --- has been simulated with different software packages to translate the scientific
requirements on the image quality into tolerances on the telescope structure and mirrors. 

The shape and size of the PSF, defined here as the 80\% containment diameter on the focal plane for the photon flux from a point source, has been simulated with
ZEMAX, with the sim\_telarray Monte Carlo programme \cite{Ber2008} used within the CTA consortium, and with the ROOT-based ROBAST ray-tracing software \cite{Oku2011}. 

\begin{figure}[h]
  \centering
\includegraphics[width=\columnwidth]{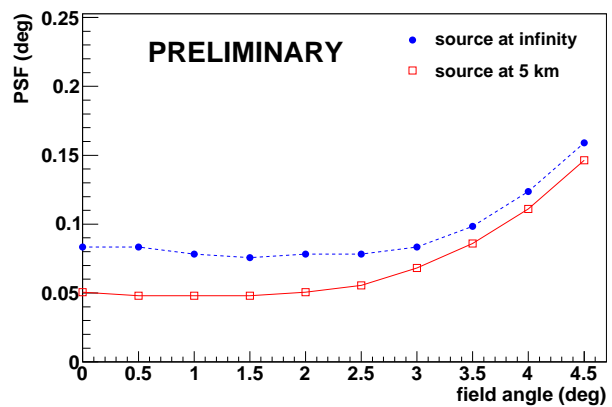}
  \caption{\small{Simulated size of the PSF as a function of the field angle for a point source at infinity and at distance of 5\,km.}}
  \label{fig:psf}
\end{figure}

As an example, the size of the PSF is shown in Fig.~\ref{fig:psf} as a function of the off-axis field angle, for point-sources at infinity and at a distance of 5\,km; the latter would correspond approximately to the distance of a very energetic air shower. It can be seen that over most of the FoV, the PSF is much smaller than the angular size of about 0.15$^\circ$ corresponding to a 6\,mm pixel. It is still contained within a single pixel at the edge of the FoV (field angle of 4.5$^\circ$). 
The results shown here are based on simulations with sim\_telarray and have been cross-checked with both ZEMAX and ROBAST. By simulating the PSF for small displacements in M1 and M2, tolerances of the mirror positions have been determined at $\sim$5\,mm for M1 and $\sim$2\,mm for M2.

Another important characteristic of the optical performance, especially for a dual-mirror telescope, is the fraction of light that is lost due to obscuration.
The backside of M2 blocks 25\% of the photons that would normally fall on an ideal M1 (without gaps and central hole) in the chosen configuration. Simulations
with sim\_telarray show that for large field angles in addition up to 2.5\% of the photon flux is absorbed by the camera body. With a detailed simulation using the 
ROBAST package, the effect of shadowing by the mast structure of the SST-GATE telescope was determined to add another 11\% to 13\% to the overall loss in throughput, depending on the field angle. It should be noted that this contribution is very similar to the one estimated for current DC telescopes. The effect of shadowing is shown
in Fig.~\ref{fig:shadowing} for an on-axis light source. The contribution from the masts that hold M2 and the arms that support the camera is clearly visible. 

\begin{figure}[!h]
  \centering
\includegraphics[width=0.8\columnwidth]{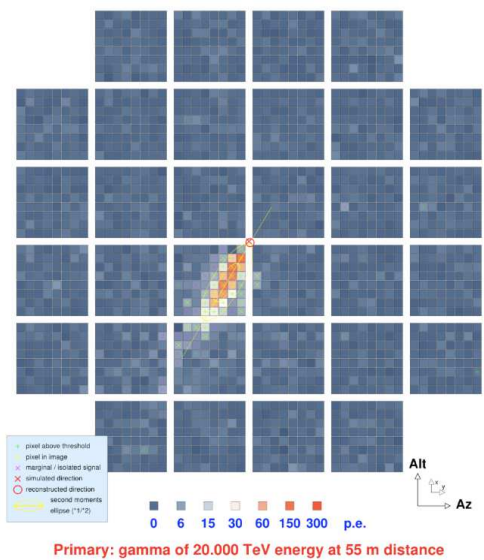}
  \caption{\small{Image of an air shower triggered by a 20\,TeV $\gamma$-ray simulated using CORSIKA and sim\_telarray (preliminary result). }}
  \label{fig:showerimage}
\end{figure}

\begin{figure}[htb]
\begin{minipage}[t]{0.49\columnwidth}
\mbox{}\\
\centerline{\includegraphics[width=\textwidth]{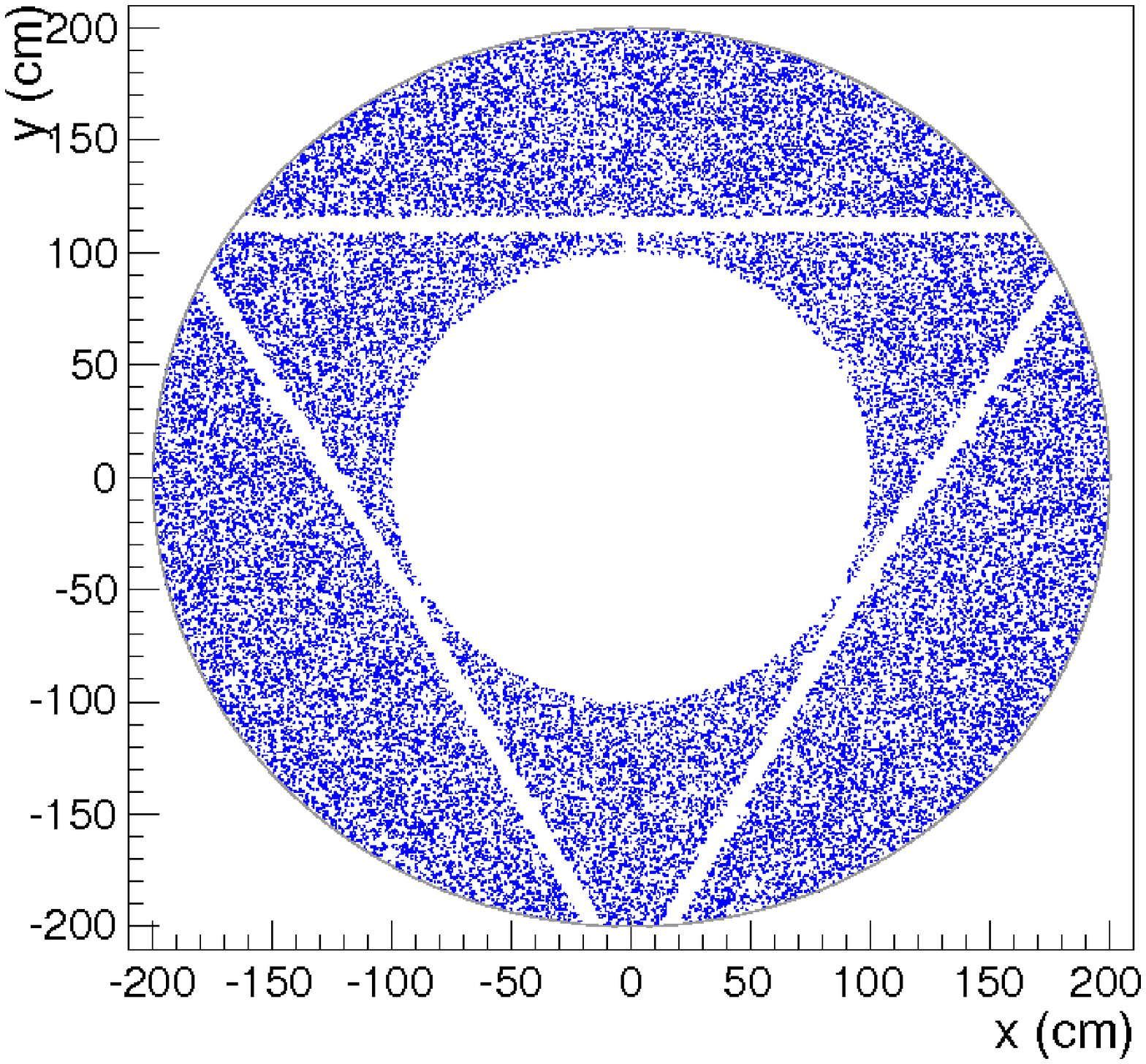}}
\end{minipage}
\hfill
\begin{minipage}[t]{0.49\columnwidth}
\mbox{}\\
\centerline{\includegraphics[width=\textwidth]{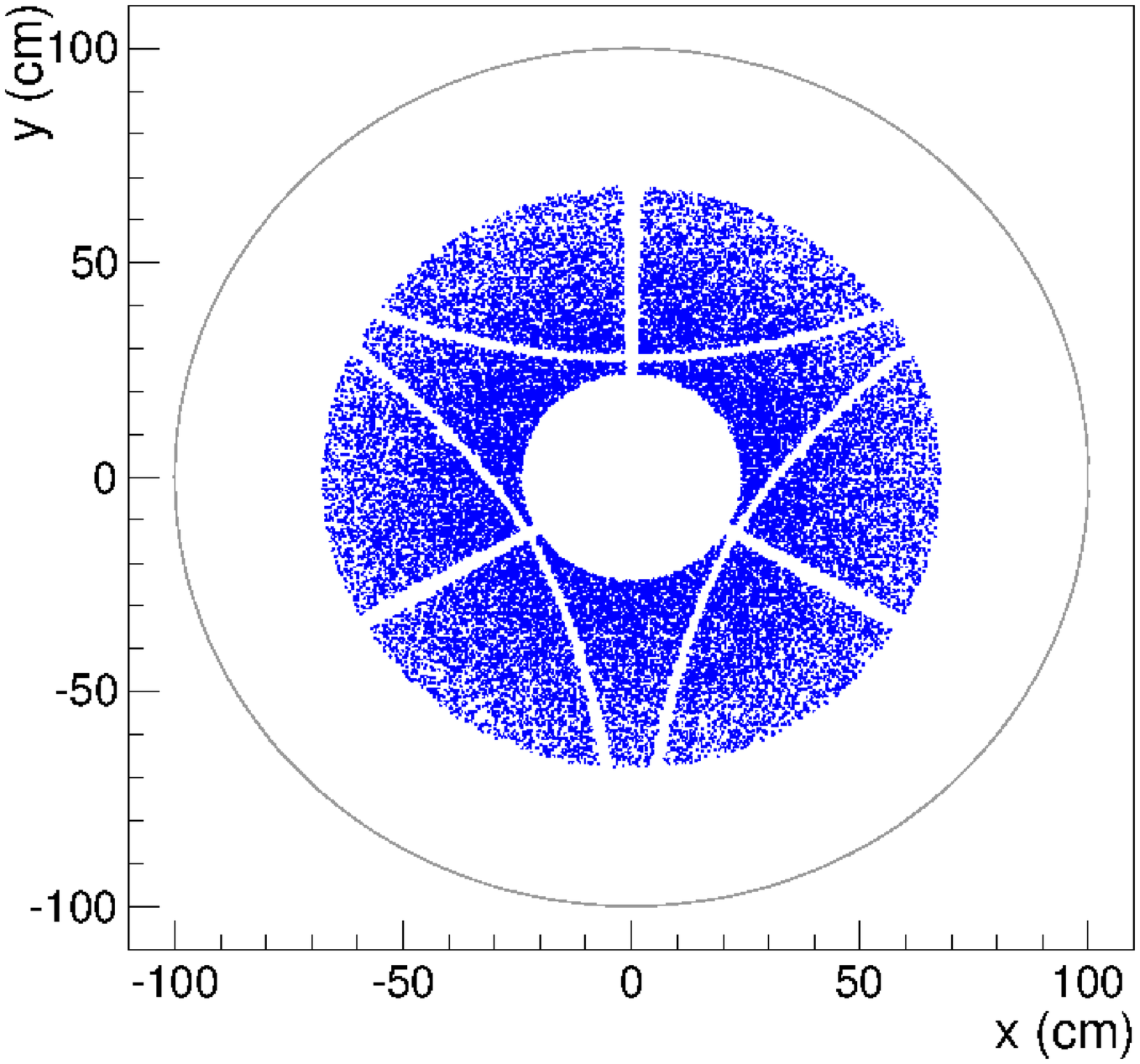}}
\end{minipage}
\caption{Simulation of light rays from an on-axis point source at a distance of 10\,km using ROBAST.  The effect of shadowing on the primary (left) and on the secondary (right) mirror becomes apparent.}
\label{fig:shadowing}
\end{figure}

Very preliminary simulations of the response of the SST-GATE optical system to simulated air showers have been made using sim\_telarray in combination with the air-shower code CORSIKA \cite{corsika}. An example for a simulated image of an air shower, triggered by a 20\,TeV $\gamma$-ray, is shown in Fig.~\ref{fig:showerimage}. 
It should be noted that these simulations do not yet include the full electronics of the CHEC camera, but they provide a good impression of the fine pixelation of the shower
image and the large FoV of the camera.

\section{Current Status and Outlook}

With the foundations, tower and shelter in place on the site of the Paris Observatory in Meudon, the first elements of SST-GATE have now been mounted. Electrical and Ethernet supplies and optical fibres are installed on the site of the prototype to provide the connection of the control software and camera to the nearby control room. The next step in the assembly process will be the mounting of the azimuth sub-system; then the remaining structure will follow. 

The fabrication procedure of M1 is currently under test and the manufacturing of the six petals will follow the test phase, so that the complete mirror could be mounted in 2014. At the same time, the mirror M2 should be received and both will be assembled on the telescope and correctly aligned in order to carry out optical and mechanical tests of the telescope. Some tests of the mechanical structure of the telescope will already be realized during construction as the software is split into dedicated sub-programmes, for instance the azimuth and elevation measure of the telescope parameters (non-verticality, non perpendicularityÉ) etc. This strategy also eases the software validation. 

An evaluation of the completed structure and software will be made in 2014. A CCD camera placed on the focal plane will allow testing of the optical quality and measuring the PSF versus field angle to compare with simulations. On delivery of the CHEC camera, the telescope will be fully mounted and final tests will be made by observing Cherenkov light from cosmic-ray induced air showers on the site in Meudon.

Upon successful completion of all tests, the different SST designs will be validated for use in the CTA. Afterwards, the SST-GATE prototype will be used as a test-bench for CTA, but also for the Master's programme of the Paris Observatory and for outreach activities. 

Within the CTA array, the SST component will be crucial for covering the highest energies. This is especially important for the observation of Galactic sources with hard spectra, such as Pulsar Wind Nebulae and Supernova Remnants, expected to extend above 100 TeV. Good energy coverage at these extreme energies will permit, for the first time, to 
study the turnover in these spectra and provide crucial information on the underlying particle distribution and the conditions inside the sources. Currently the integration of SSTs is only foreseen at the CTA site in the Southern Hemisphere. However, the addition of SSTs to the Northern Hemisphere site may also be very useful, since some of the brightest blazars seen in the Northern Sky, notably Mrk\,421 and Mrk\,501, have hard spectra that extend to at least a few 10s of TeV.

It is important to point out that, thanks to its large SST sub-array, the CTA will not only improve vastly on the performance of current IACT experiments, but will also extend once more the upper limit of the observable electromagnetic energy spectrum. 

\vspace*{0.5cm}
\footnotesize{{\bf Acknowledgment:}{ Results shown in Fig. 3 and 4 are based on O. Hervet's Master Thesis. This work has been funded under the Convention 10022639 between the R\'{e}gion d'Ile-de-France and the Observatoire de Paris. We gratefully acknowledge the Ile-de-France, the CNRS (INSU and IN2P3), the CEA and the Observatoire de Paris for financial and technical support. We also acknowledge support from the agencies and organizations listed in this page:Êhttp://www.cta-observatory.org/?q=node/22}}

\footnotesize{ 
 
}

\end{document}